# Dynamic User Interface Generation for Enhanced Human-Computer Interaction Using Variational Autoencoders

1st Runsheng Zhang
University of Southern California
Los Angeles, USA

2nd Shixiao Wang
School of Visual Arts
New York, USA

3rd Tianfang Xie
Georgia Institute of Technology
Atlanta, USA

4th Shiyu Duan
Carnegie Mellon University
Pittsburgh, USA

5th Mengmeng Chen*
New York University
New York, USA

*Abstract*—This study presents a novel approach for intelligent user interaction interface generation and optimization, grounded in the variational autoencoder (VAE) model. With the rapid advancement of intelligent technologies, traditional interface design methods struggle to meet the evolving demands for diversity and personalization, often lacking flexibility in real-time adjustments to enhance user experience. Human-Computer Interaction (HCI) plays a critical role in addressing these challenges, as it focuses on creating interfaces that are not only functional but also intuitive and responsive to user needs. This research leverages the RICO dataset to train the VAE model, enabling the simulation and creation of user interfaces that align with user aesthetics and interaction habits. Through the integration of real-time user behavior data, the system dynamically refines and optimizes the interface to improve usability, thereby underscoring the importance of HCI in achieving a seamless user experience. The experimental findings indicate that the VAE-based approach significantly enhances the quality and precision of interface generation compared to other methods, including autoencoders (AE), generative adversarial networks (GAN), conditional generative adversarial networks (cGAN), deep belief networks (DBN), and variational autoencoder generative adversarial networks (VAE-GAN). This work contributes valuable insights into Human-Computer Interaction (HCI), providing robust technical solutions for automated interface generation and enhanced user experience optimization.

I. INTRODUCTION

In recent years, with the continuous advancement of artificial intelligence technology, user interaction interface generation has gradually become a key area of research and application. The design of the interaction interface directly affects the user experience. A high-quality interaction interface can make users operate more smoothly and obtain information more efficiently. However, traditional interaction interface design often requires experienced designers to perform manual design, which is time-consuming, labor-intensive, and difficult to optimize based on real-time user feedback. Based on this, the automatic generation of interaction interfaces has become a trend. Among them, variational autoencoders (VAEs), as a generative deep learning model [1], learn user preferences through automatic encoding and decoding and achieve highly intelligent interaction interface generation, providing a new method for interaction design [2].

Variational autoencoders are highly flexible and adaptable when generating user interaction interfaces [3]. First, VAE can capture the diverse needs of users and generate interfaces of different styles and layouts by learning a large amount of user data and interface styles to meet the preferences of different user groups [4]. This adaptability is particularly critical in user interfaces because each user's needs and operating habits may be completely different, and traditional methods are difficult to meet personalized needs. Through VAE technology, the system can not only automatically generate an interface that meets the user's aesthetic taste but also optimize it according to the actual application scenarios and interaction habits, ultimately providing users with a more personalized and intuitive operation experience [4].

Secondly, the interactive interface generation process based on variational autoencoders can achieve a high degree of efficiency [5]. Compared with manual design, the efficiency of automatic generation is undoubtedly higher, and VAE, through its powerful generation ability, can generate a large number of different interface samples in a very short time and optimize the design details in real time. This not only greatly shortens the interface design cycle but also reduces the design cost, allowing designers to focus more on innovative design and complex function implementation [6]. At the same time, the interface generated by VAE is of reliable quality, can maintain the unity and logic of the design, and provides technical support for large-scale, fast-iteration projects.

Variational autoencoder (VAE) technology has significant potential in interface optimization. Unlike traditional designs requiring user testing and manual adjustments, VAEs autonomously learn user behavior patterns and adjust interface layouts and settings, enhancing user experience through dynamic, intelligent updates. By iteratively refining interfaces based on feedback, VAEs achieve a high degree of alignment with user needs, eliminating the need for manual modifications. VAEs are also highly scalable and versatile. They can integrate with models like natural language processing (NLP) [7]to generate interface elements based on user input, helping users locate functions more efficiently [8]. VAEs are crucial in diverse fields. In financial risk analysis [9], they detect anomalies in high-dimensional data [10]and synthesize plausible scenarios for stress-testing [11]. In computer vision [12], VAEs aid in image generation, anomaly detection, and data reconstruction [13], addressing challenges in areas like medical imaging [14]and classification [15-16]. Their modular design enables seamless application across platforms, such as web and mobile, meeting diverse user requirements. The use of VAEs in interface design lays the groundwork for advanced human-computer interaction. As technology evolves, VAEs will enable interfaces to predict user needs and adapt in real-time, promoting more personalized and seamless interactions while advancing the field of intelligent interfaces.

## II. METHOD

The method of this study is based on the core idea of the variational autoencoder (VAE) model [17], which aims to achieve efficient data generation and optimization through a probabilistic generation model. The network architecture of VAE is shown in Figure 1.

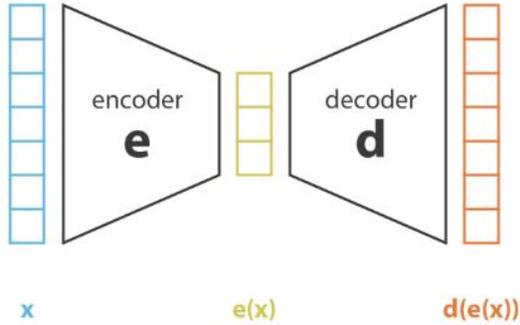

Figure 1 VAE Network Architecture

VAE mainly optimizes the generation model by maximizing the log-likelihood probability $P(x)$ of the observed data [18]. Let $x$ be the interface data sample, and the goal is to learn the implicit representation $z$ from it so that the model can generate an interface similar to $x$ under the condition of given $z$ [19]. First, consider the log-likelihood of the observed sample $x$:

$$\log p(x) = \int q(z|x) \log \frac{p(x,z)}{q(z,x)} dz$$

Among them, $q(x|z)$ is the variational distribution, which is used to approximate the posterior distribution $p(z|x)$. By introducing the variational distribution $q(x|z)$, the above formula can be further decomposed into two parts:

$$\log p(x) = E_{q(z|x)}[\log p(x|z)] - D_{KL}(q(z|x) \| p(z))$$

Among them, the first term is the reconstruction error [20], which represents the probability of generating data given the latent variable $z$; the second term is the Kullback-Leibler divergence [21], which is used to measure the distance between the approximate distribution and the true prior distribution $p(z)$. By minimizing the sum of the reconstruction error and the KL divergence, the interface generation model can be optimized [22].

The encoder consists of three fully connected layers with 512, 256, and 128 neurons, respectively, followed by ReLU activation functions. The latent space has a dimensionality of 64. The decoder mirrors the encoder with layers of 128, 256, and 512 neurons, respectively, using ReLU activations. The output layer of the decoder applies a sigmoid activation function to generate normalized interface layouts.

In order to achieve the continuity and diversity of interface generation, this study further introduced the Gaussian distribution assumption of the latent variable $z$ [23]. Assuming that the prior distribution of the latent variable $z$ is the standard normal distribution $p(z) = N(0, I)$, its conditional distribution $p(x)$ can also be assumed to be a Gaussian distribution, that is:

$$q(z|x) = N(\mu(x), \sigma^2(x))$$

Where $\mu(x)$ and $\sigma(x)$ are parameterized network outputs. During model training, the reparameterization trick is used to ensure the transferability of the gradient. The latent variable $z$ is expressed as:

$$z = \mu(x) + \sigma(x) \cdot \varepsilon$$

Where $\varepsilon \sim N(0, I)$ is the standard normal distribution noise. Through this transformation, noise can be injected into the latent variable generation process to ensure the diversity of the generated interface.

During the optimization process, the objective function is the sum of the reconstruction error and the KL divergence:

$$L = E_{q(z|x)}[\log p(x|z)] - D_{KL}(q(z|x) \| p(z))$$

The model parameters are updated by the gradient descent method so that the generated interface can better meet the personalized needs of users while maintaining a certain degree of authenticity.

In order to further realize the adaptive layout of interface elements, this study introduces a feedback mechanism based on user behavior. The user's behavior data in the interactive interface can be used as an additional input to the model, thereby affecting the generation and adjustment of the interface. Assuming that the user feedback is $f$, the generated interface can be expressed as $p(x|z,f)$. By modeling this conditional distribution, the interface can be dynamically optimized according to the user's operating habits and preferences. The final model objective function is expressed as:

$$L = E_{q(z|x)}[\log p(x|z,f)] - D_{KL}(q(z|x) \| p(z))$$

This approach draws inspiration from the work of Q. Sun et al [24]. Their research highlights the use of advanced CNN techniques for gesture recognition, which inform the dynamic and adaptive aspects of UI design in this study. By leveraging similar principles, the proposed method aims to enhance user interaction through personalized and behavior-responsive interface layouts.

In summary, this research method constructs a user interaction interface generation mechanism through a variational autoencoder model and optimizes the interface by combining user feedback. The model can not only generate a variety of interfaces based on sample data but also adjust the layout and style of the generated interface through user behavior data during the optimization process, providing users with a more personalized interaction experience that meets their needs.

III. EXPERIMENT

A. Datasets and Settings

For the experimental part of the study, this paper selected the RICO dataset. This dataset was jointly created by multiple research institutions such as MIT and the University of Pennsylvania and is currently one of the most widely used user interface (UI) datasets. The RICO dataset contains more than 90,000 screenshots of user interfaces of mobile applications, covering a variety of interface elements, including buttons, text boxes, icons, navigation bars, and other UI components. In addition, it also provides interactive layout and user behavior information to support more in-depth UI analysis and generation tasks. The application categories in the dataset are diverse, including common fields such as social, education, entertainment, and health, providing researchers with a wide range of interface samples to facilitate the exploration and learning of user interface design patterns in different categories of applications. In this study, we utilized a subset of the RICO dataset comprising 80,000 samples for training, 5,000 for validation, and 5,000 for testing to ensure robust model evaluation and prevent overfitting. The VAE model was trained using a batch size of 64 and for 200 epochs. The learning rate was initially set to 0.001 and adjusted dynamically based on validation loss. The AdamW optimizer was used to enhance generalization. All experiments were conducted on a system equipped with an NVIDIA RTX 3090 GPU, 128GB RAM, and a 12-core Intel Xeon processor running Ubuntu 20.04 and PyTorch 1.11.

B. Experimental Results

In the experiment, in order to comprehensively evaluate the performance of the variational autoencoder (VAE) model in user interface generation and optimization, this study selected five representative comparison models. These models include the traditional autoencoder (AE), which can perform basic reconstruction of data; the generative adversarial network (GAN) [25], which has strong generative adversarial properties and is suitable for high-quality image generation; the conditional generative adversarial network (cGAN), which can generate interfaces of a specific style according to specific conditions; the deep belief network (DBN) [26], which is good at feature learning and multi-layer generation tasks; and the variational autoencoder-generative adversarial network (VAE-GAN), which combines the advantages of VAE and GAN and has strong generation capabilities. These models cover from classic to advanced generation methods, providing a multi-angle reference for evaluating the effectiveness of the VAE model in UI generation.

Structural similarity (SSIM) and mean absolute error (MAE) are the primary evaluation metrics. SSIM measures the similarity between the generated and real interfaces, with higher values indicating better generation quality. MAE evaluates the average numerical error between the generated and target interfaces, reflecting the model's accuracy. Together, these metrics offer a clear basis for quantitatively comparing model performance in interface generation.

Table 1 Experimental Results

| Model | SSIM | MAE |
|---|---|---|
| AE | 0.72 | 0.130 |
| GAN | 0.78 | 0.110 |
| CGAN | 0.81 | 0.095 |
| DBN | 0.75 | 0.120 |
| VAE-GAN | 0.85 | 0.091 |
| VAE(Ours) | 0.89 | 0.073 |

Table 1 shows that the variational autoencoder (VAE) outperforms other models in user interface generation tasks, achieving an SSIM of 0.89 and an MAE of 0.073. This indicates that VAE generates interfaces with high similarity to real data and minimal reconstruction error. In comparison, GAN and VAE-GAN show lower SSIMs of 0.78 and 0.85 and higher MAEs of 0.110 and 0.091, respectively, highlighting VAE's superior stability and accuracy. VAE's combination of reconstruction loss and KL divergence ensures stable training, while its latent variable learning enables diverse and personalized interface generation. These results demonstrate VAE's effectiveness and potential for advancing automated interface design and optimization. Then we also give the loss function decline graph during training, as shown in Figure 2.

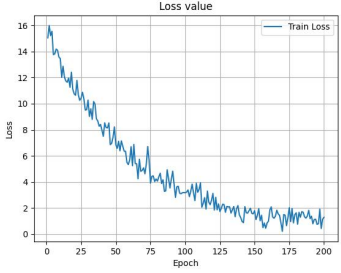

Figure 2 Loss function drop graph

Figure 2 shows a steady decrease in training loss as iterations progress, reflecting the model's continuous learning and optimization. The initial high loss, caused by unadjusted parameters, quickly declines as the model captures intrinsic data features, improving the alignment between generated and real interfaces. In later stages, the loss stabilizes, indicating effective convergence where reconstruction error and KL divergence are balanced. This balance ensures generated interfaces achieve both diversity and consistency. The consistent loss reduction without overfitting highlights the model's robust learning and generalization capabilities. Hyperparameter sensitivity experiments, starting with learning rate analysis, further evaluate its impact on training performance.

Table 2 The impact of different learning rates on experimental results

| Lr | SSIM | MAE |
|---|---|---|
| Lr=0.005 | 0.82 | 0.083 |
| Lr=0.003 | 0.84 | 0.079 |
| Lr=0.002 | 0.86 | 0.075 |
| Lr=0.001 | 0.89 | 0.073 |

The results in Table 2 demonstrate that learning rates significantly influence the model's performance in interface generation tasks. As the learning rate decreases, the model shows consistent improvements in structural similarity (SSIM) and mean absolute error (MAE). For instance, at a learning rate of 0.005, the model achieves an SSIM of 0.82 and an MAE of 0.083, while reducing the learning rate to 0.001 improves SSIM to 0.89 and lowers MAE to 0.073. The learning rate, a critical hyperparameter, determines the step size for parameter updates during training. A large learning rate accelerates parameter updates but can destabilize convergence near the global optimum, degrading output quality. For example, with a learning rate of 0.005, the model generates less realistic interfaces, likely due to insufficient stabilization during training. Gradually lowering the learning rate to 0.003 and 0.002 improves SSIM to 0.84 and 0.86 and reduces MAE to 0.079 and 0.075, as more stable parameter updates allow the model to capture finer data features and generate more accurate interfaces.

At the optimal learning rate of 0.001, the model achieves its best results, with an SSIM of 0.89 and an MAE of 0.073. A smaller learning rate facilitates fine-tuning near the global optimum, improving accuracy and consistency while minimizing risks of overfitting or divergence. This setting allows the model to balance reconstruction error and KL divergence effectively, enhancing interface generation quality and generalizability. In summary, the findings underscore the importance of learning rate selection in optimizing VAE performance for interface generation. Reducing the learning rate improves the balance between diversity and consistency, leading to interfaces that better align with user expectations. These insights provide a valuable reference for future research, emphasizing the role of careful learning rate adjustment in complex generative tasks. Additionally, this study examines the influence of different optimizers, as detailed in Table 3.

Table 3 The impact of different optimizers on experimental results

| Lr | SSIM | MAE |
|---|---|---|
| RMSprop | 0.81 | 0.084 |
| Adam | 0.83 | 0.080 |
| SGD | 0.85 | 0.077 |
| AdamW | 0.89 | 0.073 |

The results in Table 3 highlight significant differences in model performance across optimizers for the user interface generation task. As the optimizer changes from RMSprop to AdamW, the model's structural similarity index measure (SSIM) improves, and mean absolute error (MAE) decreases, indicating that selecting an appropriate optimizer enhances interface generation quality. Specifically, RMSprop achieves an SSIM of 0.81 and an MAE of 0.084, while AdamW achieves the best results with an SSIM of 0.89 and an MAE of 0.073, demonstrating its superior performance.

RMSprop, which adapts the learning rate using the sliding average of squared gradients, performs adequately but shows limited effectiveness in refining interface details. In contrast, the Adam optimizer, with an SSIM of 0.83 and an MAE of 0.080, outperforms RMSprop by leveraging momentum and adaptive learning rate adjustments, improving stability and robustness in gradient updates. The SGD optimizer further enhances performance, achieving an SSIM of 0.85 and an MAE of 0.077. Although less efficient than adaptive optimizers, SGD's inherent noise in parameter updates aids regularization and detail capture, reducing overfitting. However, its performance remains constrained by its reliance on manual learning rate adjustments.

The AdamW optimizer delivers the best results, combining the advantages of Adam with weight decay to improve generalization and optimization quality. This approach enhances model performance in complex tasks, producing interfaces that are closer to real samples and exhibit higher detail precision. Overall, AdamW emerges as the most effective optimizer for this study, making it a preferred choice for user interface generation tasks.

IV. CONCLUSION

The user interaction interface generation and optimization method proposed in this study highlights the significance of Human-Computer Interaction (HCI) in creating adaptive, intelligent, and user-centered interfaces. Traditional interface design struggles to accommodate the dynamic and personalized needs of users, which is where HCI plays a transformative role. By focusing on user needs and behaviors, this approach supports the design of interfaces that are highly intuitive, engaging, and responsive.

Through extensive use of the RICO dataset, the proposed method demonstrates a strong capability to learn implicit features of diverse interface designs, enabling it to adapt dynamically to different user requirements. The system leverages real-time feedback from user behavior data to refine and optimize interfaces, showcasing the potential of HCI in elevating both usability and user satisfaction. The experimental results illustrate the model's effectiveness in maintaining interface quality and consistency while enhancing the personalization of user interactions. The intelligent optimization process presented in this study represents an important advancement in HCI. By dynamically updating interfaces during actual use and adaptively adjusting layouts and features in response to user behavior, the system can provide users with an improved and more personalized interactive experience. This real-time optimization mechanism is a core aspect of HCI, emphasizing the importance of adaptive design that aligns with users' operating habits, thereby enhancing user satisfaction and engagement.

Future research should continue to explore ways to enhance HCI through the integration of various deep learning models and technologies. For example, combining the current approach with natural language processing (NLP) could enable the system to generate more context-aware interface elements, further facilitating the user's ability to find required functions efficiently. Another avenue for future work is to investigate the fusion of multimodal data—such as text inputs, user behavior, and image preferences—enabling a more comprehensive understanding of user needs and improving the adaptability of interface generation. In summary, this study underscores the critical role of HCI in intelligent interface generation and optimization, providing new pathways for automated and user-centered design. As user needs become increasingly diverse and intelligent, the emphasis on HCI will be crucial in driving forward innovations that offer personalized, real-time, and adaptive interaction experiences, ultimately contributing to a higher level of user experience and satisfaction.

REFERENCES

[1] J. Hu, Y. Cang, G. Liu, M. Wang, W. He, and R. Bao, "Deep Learning for Medical Text Processing: BERT Model Fine-Tuning and Comparative Study", arXiv preprint arXiv:2410.20792, 2024.

[2] A. Terović and I. Mekterović, "On Improving the Qualitative Features of the User Interface of Mobile Applications Using Machine Learning Methods," Proceedings of the 2024 47th MIPRO ICT and Electronics Convention (MIPRO), pp. 205-210, 2024.

[3] L. Chen, Q. Jing, Y. Tsang, et al., "Iris: a multi-constraint graphic layout generation system," Frontiers of Information Technology & Electronic Engineering, vol. 25, no. 7, pp. 968-987, 2024.

[4] L. Chen, Q. Jing, Y. Zhou, et al., "Element-conditioned GAN for graphic layout generation," Neurocomputing, vol. 591, p. 127730, 2024.

[5] Y. Feng, A. Shen, J. Hu, Y. Liang, S. Wang, and J. Du, "Enhancing Few-Shot Learning with Integrated Data and GAN Model Approaches", arXiv preprint arXiv:2411.16567, 2024.

[6] D. Tamayo-Urgilés, S. Sanchez-Gordon and Á. L. Valdivieso Caraguay, "GAN-Based Generation of Synthetic Data for Vehicle Driving Events," Applied Sciences, vol. 14, no. 20, p. 9269, 2024.

[7] C. Tao, X. Fan, and Y. Yang, "Harnessing LLMs for API Interactions: A Framework for Classification and Synthetic Data Generation," arXiv preprint arXiv:2409.11703, 2024.

[8] J. Chen, B. Liu, X. Liao, J. Gao, H. Zheng, and Y. Li, "Adaptive Optimization for Enhanced Efficiency in Large-Scale Language Model Training," arXiv preprint, 2024.

[9] W. Sun, Z. Xu, W. Zhang, K. Ma, Y. Wu, and M. Sun, "Advanced Risk Prediction and Stability Assessment of Banks Using Time Series Transformer Models", arXiv preprint arXiv:2412.03606, 2024.

[10] Z. Xu, W. Zhang, Y. Sun, and Z. Lin, "Multi-Source Data-Driven LSTM Framework for Enhanced Stock Price Prediction and Volatility Analysis", Journal of Computer Technology and Software, vol. 3, no. 8, 2024.

[11] Z. Liu, X. Xia, H. Zhang and Z. Xie, "Analyze the Impact of the Epidemic on New York Taxis by Machine Learning Algorithms and Recommendations for Optimal Prediction Algorithms," Proceedings of the 2021 3rd International Conference on Robotics Systems and Automation Engineering, pp. 46-52, May 2021.

[12] S. Lu, Z. Liu, T. Liu and W. Zhou, "Scaling-up Medical Vision-and-Language Representation Learning with Federated Learning," Engineering Applications of Artificial Intelligence, vol. 126, Article ID 107037, 2023.

[13] A. Shen, M. Dai, J. Hu, Y. Liang, S. Wang, and J. Du, "Leveraging Semi-Supervised Learning to Enhance Data Mining for Image Classification under Limited Labeled Data", arXiv preprint arXiv:2411.18622, 2024.

[14] J. Song and Z. Liu, "Comparison of Norm-Based Feature Selection Methods on Biological Omics Data," Proceedings of the 5th International Conference on Advances in Image Processing, pp. 109-112, November 2021.

[15] Y. Xiao, "Self-Supervised Learning in Deep Networks: A Pathway to Robust Few-Shot Classification", arXiv preprint arXiv:2411.12151, 2024.

[16] B. Chen, F. Qin, Y. Shao, J. Cao, Y. Peng and R. Ge, "Fine-Grained Imbalanced Leukocyte Classification With Global-Local Attention Transformer," Journal of King Saud University - Computer and Information Sciences, vol. 35, no. 8, Article ID 101661, 2023.

[17] Y. Yao, "Self-Supervised Credit Scoring with Masked Autoencoders: Addressing Data Gaps and Noise Robustly", Journal of Computer Technology and Software, vol. 3, no. 8, 2024.

[18] Y. Yang, I. Li, N. Sang, L. Liu, X. Tang, and Q. Tian, "Research on Large Scene Adaptive Feature Extraction Based on Deep Learning", Preprints, doi: 10.20944/preprints202409.0841.v1, 2024.

[19] J. Wei, Y. Liu, X. Huang, X. Zhang, W. Liu and X. Yan, "Self-Supervised Graph Neural Networks for Enhanced Feature Extraction in Heterogeneous Information Networks", 2024 5th International Conference on Machine Learning and Computer Application (ICMLCA), pp. 272-276, 2024.

[20] J. Cao, R. Xu, X. Lin, F. Qin, Y. Peng and Y. Shao, "Adaptive Receptive Field U-Shaped Temporal Convolutional Network for Vulgar Action Segmentation," Neural Computing and Applications, vol. 35, no. 13, pp. 9593-9606, 2023.

[21] J. Du, G. Liu, J. Gao, X. Liao, J. Hu, and L. Wu, "Graph Neural Network-Based Entity Extraction and Relationship Reasoning in Complex Knowledge Graphs", arXiv preprint arXiv:2411.15195, 2024.

[22] X. Wang, X. Li, L. Wang, T. Ruan, and P. Li, "Adaptive Cache Management for Complex Storage Systems Using CNN-LSTM-Based Spatiotemporal Prediction", arXiv preprint arXiv:2411.12161, 2024.

[23] W. Liu, Z. Zhang, X. Li, J. Hu, Y. Luo, and J. Du, "Enhancing Recommendation Systems with GNNs and Addressing Over-Smoothing", arXiv preprint arXiv:2412.03097, 2024.

[24] Q. Sun, T. Zhang, S. Gao, L. Yang, and F. Shao, "Optimizing Gesture Recognition for Seamless UI Interaction Using Convolutional Neural Networks", arXiv preprint arXiv:2411.15598, 2024.

[25] Prabhat, Nishant and D. Kumar Vishwakarma, "Comparative Analysis of Deep Convolutional Generative Adversarial Network and Conditional Generative Adversarial Network using Hand Written Digits," 2020 4th International Conference on Intelligent Computing and Control Systems (ICICCS), pp. 1072-1075, 2020.

[26] R. Gupta, S. Kumar Shukla and V. Tripathi, "New Deep Learning Models for Medical Imaging: Deep Belief Network, GAN,

Autoencoder," 2023 4th International Conference on Smart Electronics and Communication (ICOSEC), Trichy, India, 2023.